\documentclass[12pt,dvips]{article}
\usepackage{graphicx}
\usepackage{amsmath}
\usepackage[psamsfonts]{amssymb}
\usepackage{amsthm}
\usepackage{euscript}

\usepackage{latexsym}

\renewcommand{\theequation}{\arabic{section}.\arabic{equation}}
\renewcommand{\(}{\begin{equation}}
\renewcommand{\)}{end{equation} \vspace{-.05in}\linebreak}

\newcounter{saveeqn}
\newcounter{savealpheqn}

\newcommand{\alpheqn}{\setcounter{saveeqn}{\value{equation}}%
 \stepcounter{saveeqn}\setcounter{equation}{0}%
 \renewcommand{\theequation}{\mbox{\arabic{section}.\arabic{saveeqn}\alph{equation}}}
 \renewcommand{\)}{\end{equation}}}
\def\part#1{\frac{\partial}{\partial{#1}}}%

\def\group#1{\refstepcounter{equation}\setcounter{saveeqn}{\value{equation}}%
\label{#1}\setcounter{equation}{0}%
\renewcommand{\theequation}{\mbox{\arabic{section}.\arabic{saveeqn}\alph{equation}}}%
\renewcommand{\)}{\end{equation}}}

\newcommand{\reseteqn}{\setcounter{equation}{\value{saveeqn}}%
 \renewcommand{\theequation}{\arabic{section}.\arabic{equation}}%
 \renewcommand{\)}{\end{equation}}}

\newcommand{\aalpheqn}{\setcounter{saveeqn}{\value{equation}}%
 \stepcounter{saveeqn}\setcounter{equation}{0}%
 \renewcommand{\theequation}{\mbox{\Alph{subsection}.\arabic{saveeqn}\alph{equation}}}
  \renewcommand{\)}{\end{equation}}}

\newcommand{\areseteqn}{\setcounter{equation}{\value{saveeqn}}%
 \renewcommand{\theequation}{\Alph{subsection}.\arabic{equation}}%
 \renewcommand{\)}{\end{equation}}}

\renewcommand{\thefootnote}{\alph{footnote}}
\renewcommand{\(}{\begin{equation}}
\renewcommand{\)}{\end{equation}}
\newcommand{\ba}{\begin{eqnarray}}
\newcommand{\ea}{\end{eqnarray}}

\newcommand{\bp}{\mathop{\vtop{\ialign{##\crcr
  $\hfil\displaystyle{}\hfil$\crcr\noalign{\kern-13pt\nointerlineskip}
  \BIG{(}\hskip0pt\crcr\noalign{\kern3pt}}}}}
\newcommand{\cbp}{\mathop{\vtop{\ialign{##\crcr
  $\hfil\displaystyle{}\hfil$\crcr\noalign{\kern-13pt\nointerlineskip}
  \BIG{)}\hskip0pt\crcr\noalign{\kern3pt}}}}}
\newcommand{\pa}{\mathop{\vtop{\ialign{##\crcr
  $\hfil\displaystyle{\oplus}\hfil$\crcr\noalign{\kern+1pt\nointerlineskip}
  \hspace{.08in}$^{\alpha=0}$\hskip6pt\crcr\noalign{\kern3pt}}}}}
\renewcommand{\sp}{,\hspace{.3in}}
\newcommand{\p}{^\prime}

\newcommand{\R}{\ensuremath{\mathbb R}}

\newcommand{\Z}{\ensuremath{\mathbb Z}}

\newcommand{\beq}{\begin{equation}}
\newcommand{\eeq}{\end{equation}}

\numberwithin{equation}{section}

\catcode`\@=11
\def\vereq#1#2{\lower3pt\vbox{\baselineskip1.5pt \lineskip1.5pt
\ialign{$\m@th#1\hfill##\hfil$\crcr#2\crcr\sim\crcr}}}
\catcode`\@=12

\makeatletter
\newcommand\tabcaption{\def\@captype{table}\caption}
\makeatother
\renewcommand{\(}{\begin{equation}}
\renewcommand{\)}{\end{equation}}

\begin{document}

\begin{titlepage}
\begin{flushright}
IFUP-TH/2003/12

hep-th/0302081
\end{flushright}

\vspace{2em}
\def\thefootnote{\fnsymbol{footnote}}

\begin{center}
{\Large  Twisted K-Theory from Monodromies}
\end{center}
\vspace{1em}

\begin{center}
Jarah Evslin\footnote{E-Mail: jarah@df.unipi.it} 
\end{center}

\begin{center}
\vspace{1em}
{\em INFN Sezione di Pisa\\
     Universita di Pisa\\
     Via Buonarroti, 2, Ed. C,\\
     56127 Pisa, Italy}\\
\end{center}

\vspace{3em}
\begin{abstract}
\noindent

\end{abstract}
RR fluxes representing different cohomology classes may correspond to the same twisted K-theory class.  We argue that such fluxes are related by monodromies, generalizing and sometimes T-dual to the familiar monodromies of a D7-brane.  A generalized ``theta angle'' is also transformed, but changes by a multiple of 2$\pi$.  As an application, NS5-brane monodromies modify the twisted K-theory classification of fluxes.  Furthermore, in the noncompact case K-theory does not distinguish flux configurations in which $dG$ is nontrivial in compactly supported cohomology.  Such fluxes are realized as the decay products of unstable D-branes that wrapped nontrivial cycles.  This is interpreted using the $E_8$ bundle formalism.  

\vfill
February 11, 2003

\end{titlepage}
\setcounter{footnote}{0} 
\renewcommand{\thefootnote}{\arabic{footnote}}

\pagebreak
\renewcommand{\thepage}{\arabic{page}}
\pagebreak 

\section{Introduction}

In theories with Chern-Simons interactions, such as the SUGRA theories that provide the classical limits of string theories, there are many distinct varieties of field strength \cite{Marolf} which exhibit varying degrees of quantization and gauge invariance.  The different field strengths of type II supergravity are distinguished by their dependence on the NS fields $B$ and $H$.  In particular in the absence of NS forms there is only one type of RR field strength, the exterior derivative of the RR gauge potentials
\begin{equation}
G_{p+1}=dC_{p}.
\end{equation}
This is the case analyzed in Ref.~\cite{MW}, where it was concluded that RR fluxes are classified by K-theory.  When the NS fluxes are reintroduced it is believed that some notion of field strength is then classified by twisted K-theory \cite{BandM,DMW}.  However few examples of fluxes classified by twisted K-groups have been analyzed, and in fact one of the most careful analyses \cite{Triples} revealed a counterexample.  It was found that the set of allowed RR charges does not even necessarily form a group, a finding which may have been expected because even classically the allowed fluxes are solutions of nonlinear equations.  When the H-field is turned off these equations become linear, and so in that case one may expect their solutions to form a group.

In the present note several examples with NS fluxes are considered in an attempt to determine what notion of RR field strength is classified by twisted K-theory and when this classification may fail.  The Maldacena, Moore and Seiberg (MMS) perspective \cite{MMS} of the K-theory classification of solitons will be used extensively.  In this perspective, the relevant K-group is constructed as the set of D$p$-branes which are anomaly free quotiented by the brane configurations that may decay via ``instantonic'' D$(p+2)$-branes.  The term ``instantonic'' was coined in Ref.~\cite{MMS} and refers to the fact that these branes may be solutions to either the Euclidean or Minkowskian equations of motion.  In the first case they therefore correspond to the unstable branes tunnelling out of existence in some quantum theory.  However the idea behind this construction does not rely upon any quantum theory\footnote{After all, D-branes may not be suitable degrees of freedom in such a theory.}.  Rather, these branes are characterized by the fact that they exist for a finite amount of time.  Therefore we will refer to them as ``mortal'' branes, thus avoiding a continued notational clash with ``D-instantons''. 

At $g_s=0$ mortal branes appear, consume the unstable branes and then disappear.  When the string coupling is finite this is a continuous process in which the unstable branes blow up via the Myers effect \cite{Myers}, sweep out a cycle which causes them to lose all of their charges and then shrink into nothing
.  The unstable D$p$-branes are trivial in twisted K-theory, and thus their decay into nothing would allow us to quotient by their associated charges and thus confirm the twisted K-theory classification.  However the D$p$-branes do not quite decay into nothing, because the mortal D$(p+2)$-brane creates $G_{6-p}$ flux which remains after all of the branes have disappeared.  This means that configurations which differ only by the inclusion of an unstable stack of D-branes are not related by any dynamical process.

This effect does not affect the classification of D-branes, because the remaining flux does not carry any D-brane charge and so may be neglected\footnote{If one does not ignore the flux, one may use it to constructed twisted K-theory with every possible twist, including the twist which yields integral cohomology.  All such charges are conserved simultaneously.  Of course only one of these groups classifies D-branes at any given time, but which group this is changes whenever one crosses an NS5-brane.}.  However when classifying RR fluxes one must determine if and when the residual flux is classified by twisted K-theory.  

We begin in Section~\ref{revsec} with a review of the classical equations of motion and different notions of field strengths in $N$=2 10d supergravities.  In particular we describe the inclusion of charges and the relevant Gauss' laws, using the brane surgery viewpoint developed in Ref.~\cite{Townsend}.  In Section~\ref{exsec} we focus on a $spin^c$ example with time-independent topology, type II string theory on $S^3\times S^5\times \R^{1,1}$.  We investigate various dynamical process involving the spontaneous creations and destructions of branes and the related monodromies which may lead to the equivalence of different field strengths.  We attempt to determine which configurations are covered by the twisted K-theory classifications and which are identified.  We find that $dG$ terms which are trivial in ordinary cohomology but nontrivial in compactly supported cohomology are not represented by any K-theory class.  To understand this from another perspective, in Section~\ref{e8sec} we outline the way in which twisted K-theory appears to arise in the $E_8$ gauge bundle formalism.  We see that the fluxes in question do not affect the topology of the $E_8$ bundles, and so do not affect the K-theory class of a configuration.  Finally in Section~\ref{ns5sec} we include dynamical processes involving NS5-branes and fundamental strings and find that the twisted K-theory classification is modified so as to include less configurations and to identify more fields.  The further identifications are caused by monodromies as one encircles an NS5-brane.  NS5-branes, roughly, divide the world regions with differing values of the $H$-field, but it appears as though one may consistently classify branes and fluxes in each of those regions by the corresponding twisted K-theory.

\section{Supergravity Review} \label{revsec}

\subsection{Fields and Charges in Type II SUGRA}

The bosonic degrees of freedom in $N=2$ 10-dimensional supergravities are the graviton, the dilaton, the NS B-field $B$ and the RR $p$-form gauge connections $C_p$ where $p$ is odd and even for type IIA and IIB supergravities respectively.  The NS and RR forms enjoy the abelian gauge symmetries
\begin{equation}
B\longrightarrow B+d\omega\sp
C_p\longrightarrow C_p+d\Lambda_{p-1}-B\wedge d\Lambda_{p-3}.
\end{equation} 
From these one may construct the gauge-invariant field strengths
\begin{equation}
H=dB\sp
G_{p+1}=dC_p+H\wedge C_{p-2}.
\end{equation}
The various field strengths are related by a self-duality which takes these Bianchi identities to equations of motion, but we will not need to make use of this fact.  For notational simplicity, we will always denote the multiplication of fluxes by a wedge product as above.  However if one wishes to apply the formulas of this note to torsion cohomology classes it should be understood that the wedge products are in fact cup products.  Our examples will not contain torsion classes and so there will be no distinction.

The gauge-invariant field strengths are gauge-invariant but need not be closed.  On the other hand, the exterior derivatives of the connections, being exterior derivatives of forms which are at least patchwise well-defined, are closed although they are not gauge-invariant.  The exterior derivative of a closed form vanishes and so we find the Bianchi identities
\begin{equation}
0=ddC_p=d(G_{p+1}-H\wedge C_{p-2})=dG_{p+1}-H\wedge G_{p-1}. \label{bianchi}
\end{equation}
We may couple the theory to charged matter by allowing the Bianchi identities to be violated by a source $j_p$
\begin{equation}
ddC_p=dG_{p+1}-H\wedge G_{p-1}=j_p.
\end{equation}
Physically $j_p$ measures D$(7-p)$-brane charge.

\begin{figure}[ht]
  \centering \includegraphics[width=5in]{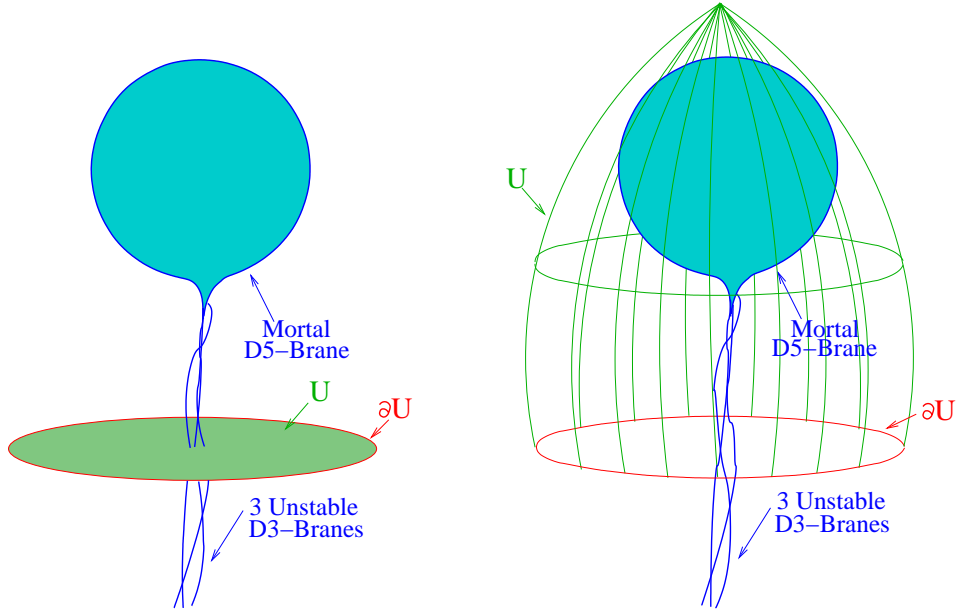}
\caption{In a spacetime with 3 units of $H$ flux, stacks of 3 D3-branes are unstable.  D3-brane charge is measured by Gauss' Law, that is by the integral of $dC_4$ over a linking 5-cycle $\partial U$.  By Stoke's Theorem this is an integral over the region $U$, but the two different choices of $U$ drawn above yield answers which differ by $3$.  This ill-definedness in D3-brane charge is a consequence of the instability of stacks of 3 D3-branes.}\label{uandu}
\end{figure}

One may calculate the total charge in a region $U$ by integrating the current $j_p$
\begin{equation}
Q_U=\int_Uj_p=\int_Ud(dC_p)=\int_{\partial U}dC_p=\int_{\partial U}G_{p+1}-H\wedge C_{p-2}.
\end{equation}
To make charges integral and equations small we have included an extra, hidden factor of $2\pi$ in each of these integrals and in all integrals that follow.  $dC_p$ is not gauge invariant and furthermore in the presence of other RR currents even its integral over $\partial U$ is not gauge invariant.  This is reflected in the nonconservation of the current $j_p$
\begin{equation}
dj_p=-H\wedge j_{p-2}.
\end{equation}
The physical interpretation is that $k$ D$(7-p)$-branes must end on any D$(9-p)$-brane that wraps a 3-cycle supporting $k$ units of $H$ flux.  In the sequel we will always consider $k\neq 0$.  The nonconservation of $j_p$ is manifested in the fact that if $\partial U$ links the D$(7-p)$-branes then the interior region $U$ may be chosen to intersect the branes, in which case the charge is detected, or it may be chosen to go around the D$(9-p)$-brane, in which case $Q_U$ vanishes.  The two choices of interior $U$ are illustrated in Figure~\ref{uandu}.

\subsection{Classification Schemes}

Thus we learn that groups of $k$ D$(7-p)$-branes can decay if a D$(9-p)$-brane bubble nucleates carrying no charges and then proceeds to absorb the D$(7-p)$-branes and sweep out a 3-cycle carrying $k$ units of $H$ flux before disappearing itself.  This implies that D$(7-p)$-charge, for each type of consistent brane, is roughly classified by the cyclic group $\Z_k$.  The RR field strength $G_{p+1}$ may be classified by a cohomology group, perhaps $H^{p+1}(M,U(1))$.  Although it is not homotopy-invariant \cite{Townsend}, there is no evidence here that $G_{p+1}$ ill-defined and so it does not appear as though a quotient needs to be taken.  It contains an integral of the $C$-field, which is not quantized, and so real or $U(1)$ coefficients would seem appropriate.  $U(1)$ coefficients are chosen here to reflect the fact that the $C_{p-2}$ is classified by $U(1)$-cohomology ($\R$ coefficients would miss, for example, discrete torsion).  

On the other hand, in the quantum theory $dC_p$ is quantized by the Dirac quantization condition, which up to gravitational corrections forces its integrals to be integral.  Its integrals are only defined modulo $k$, as evidenced by the two representatives of $U$ in Fig.~\ref{uandu} that give answers which differ by $k$.  Thus each ``type'' of $dC_p$ appears to be classified by $Z_k$.  This is in accord with the twisted K-theory classification, in fact even the gravitational corrections are naturally interpreted in this framework \cite{MW}.

However the twisted K-theory classification is not supposed to classify only a single RR flux at a time, it is conjectured to classify the entire collection of fluxes at once.  All of them are supposed to be the characteristic classes associated with a single K-theory class.  In particular, the above decay left a residual $G_{p-1}$ flux.  This remaining flux is particularly dangerous because its source wraps a cycle which supports $H$ flux, which may indicate that $G_{p-1}\wedge H$ does not vanish and may even be nontorsion.  In this case $G_{p-1}$ could not be one of the RR forms associated with twisted K-theory, whose wedge products with $H$ are always pure torsion.  However this would not violate the Bianchi identity (\ref{bianchi}), it would merely imply that $dG_{p+1}=H\wedge G_{p-1}$ is a nontorsion class\footnote{$dG_{p+1}$ may be nontrivial only as an element of cohomology with compact support.}.  As the Bianchi idenity is not violated, there is no D-brane charge and so this is a situation in which RR fields are conjectured to be classified by twisted K-theory.

Reversing the logic, if the twisted K-theory classification is to be sensitive to this RR flux, then $H\wedge G_{p-1}$ must be pure torsion at the end of the decay described above, in fact it must equal $Sq^3 G_{p-1}$.  We will now test whether this is indeed the case in examples.

\section{Examples} \label{exsec}

\subsection{The MMS Example}
\begin{figure}[ht]
  \centering \includegraphics[width=3in]{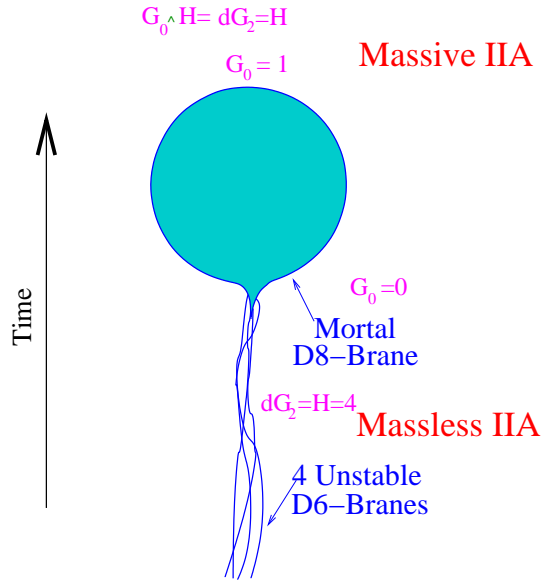}
\caption{Consider massless IIA SUGRA on a spacetime with 4 units of $H$ flux, stacks of 4 D6-branes are unstable.  The decay occurs via a mortal D8-brane.  As the D8-brane sweeps out space, it leaves behind it massive IIA with a Romans mass $G_0=1$.  This does not correspond to a twisted K-theory class because $H\wedge G_0\neq 0$.} \label{mms}
\end{figure}

Consider massless type IIA string theory on $S^3\times S^5\times \R^{1,1}$, although all of the remarks that will be made apply equally to the MMS example $SU(3)\times \R^{1,1}$.  Let there be $k$ units of $H$ flux supported on the 3-sphere.  Consider the following decay process \cite{MMS}.  $k$ static D6-branes wrap the 5-cycle and also extend along the noncompact direction.  A D8-brane forms which also wraps the 5-cycle and extends in the noncompact direction, and with its other 3 directions it sweeps out the 3-sphere.  The 3-sphere may be swept out all at one instant, making the D8-brane an S-brane \cite{S-branes}, or for example it may start as a small 2-sphere on the 3-sphere and as time progresses it may grow until it wraps an equator and then shrink out of existence on the other side.  Sometime during its finite life, let the D8-brane swallow the D6's.  If it fails, that is inconsequential as when it shrinks to a point on the 3-sphere it will have $k$ units of $\overline{\textup{D}6}$ charge and can annihilate the D6's later.  At finite $g_s$ these two processes are indistinguishable as the stack D6-branes will always be slightly dielectric \cite{HanWit} and thus will always be a spherical D8 brane.  In the end there are no branes, and so the remaining fluxes should be described by twisted K-theory.  This process is drawn in Figure~\ref{mms}.

However during the decay the world was engulfed by a bubble of D8-brane, and as usual the inside of this bubble is Romans massive IIA with a Romans mass $G_0=1$.  This flux is precisely of the dreaded form, 
\begin{equation}
dG_2=G_0\wedge H=H
\end{equation}
which is nonvanishing and nontorsion.  Stated another way
\begin{equation}
d_3 G_0=(Sq^3+H\wedge)G_0=H\neq 0
\end{equation}
where $d_3$ is a differential from the Atiyah-Hirzebruch spectral sequence.  This means that $G_0$ does not correspond to any K-theory class.

The initial state contained D-branes and so its fluxes were not supposed to be classified by twisted K-theory.  However, from this state we may have known that the final product also would not correspond to a K-theory class.  Recall that the RR fluxes are gauge invariant, and so integrals of $G_2$ and therefore $dG_2$ are well-defined over any given surface, although integrals of $G_2$ may change if the surface is moved through a D6-brane.  The initial state contained $dG_2$ flux supported on the D6-branes.  This is clear from inserting the initial condition $G_0=0$ of massless IIA into the modified Bianchi identity
\begin{equation}
dG_2-G_0\wedge H=j_1
\end{equation}
where $j_1$ is the D6-brane current.  $dG_2$ is conserved and so it is also nonvanishing in the final state, which implies that $G_0$ is not annihilated by $d_3$ and thus the final state does not represent a K-theory class.  Conversely, twisted K-theory does seem to classify field configurations which are the decay products of brane configurations in which each $dG_{p+1}$ is cohomologically trivial.

One way out of this restriction on the applicability of twisted K-theory would be to observe that $dG_{p+1}$ and in fact $H$ itself are nontrivial only as elements of cohomology of compact support.  Thus if one is interested in K-theory without compact support, one may be inclined to consider $dG_2$ and therefore $H$ to vanish.  However, with $H$ ignored (gauged away using a $B$ without compact support, although possibly with rapid decrease depending on details of the SUGRA solution) it is hard to justify the use of {\it{twisted}} K-theory in this example, as $H$ is identified with zero.  But perhaps there is some sense in which $dG_2$ is trivial but not $H$.

To restate the issue, we know from the Bianchi identity that in the absence of branes
\begin{equation}
H\wedge G_{p-1}=dG_{p+1}
\end{equation}
and so $H\wedge G_{p-1}$ may only be nontrivial as an element of compactly supported cohomology.  $H\wedge G_{p-1}$ is required to be torsion\footnote{More precisely, it must equal $Sq^3 G_{p-1}$.} in order for the field configuration to correspond to a K-theory class.  In cohomology without compact support $dG_{p+1}$ and therefore $H\wedge G_{p-1}$ vanish and so configurations do correspond to K-theory classes, but not twisted K-theory classes unless $H$ represents a class in the theory without compact support.  On the other hand $H\wedge G_{p-1}$ is often nontrivial and nontorsion in compactly supported cohomology, and so such states do not fit into the twisted K-theory classification. 


If the physics were to somehow force $dG_{p+1}$ to be trivial, then the field configurations missed by twisted K-theory classification would be eliminated.  One way to do this would be to restrict attention to compact manifolds.  $dG_{p+1}$ would then be forced to vanish to assure that $G_{p+1}$ is well defined, or equivalently $dG_{p+1}$ would vanish because it is an element of quotient of cohomology with compact support by the usual cohomology \cite{HW}, but in the compact case everything has compact support and so the quotient is trivial.  More intuitively (but less precisely), if we require spacetime to be compact then we cannot introduce D-brane charge, apparently even D-brane charge which is trivial in K-theory but not in homology, and so there will be no troublesome decay products in the final state.  Of course D-brane charge which is trivial in K-theory may form spontaneously, but $dG_{p+1}$ will be trivial throughout this process, the Bianchi identity violation in such a case comes entirely from the $H\wedge G_{p-1}$ term and this conveniently disappears when the unstable branes vanish.  

\subsection{Lower-Dimensional Branes and Monodromies}

The last paragraph supports the claim that fluxes on compact spaces are in fact classified by twisted K-theory.  A remaining question is whether the configurations on noncompact spaces which have fluxes not corresponding to any K-theory class may be pathological.  In particular the example above had only ``one noncompact direction'' and the D6-branes were extended in that direction, meaning that as one continues in that direction it is possible that the $G_2$ field strength grows ever larger.  This effect is not as severe as one may think, as Einstein's equations demand that the noncompact directions grow at the same time, and so perhaps divergences are avoided in an actual SUGRA solution.  However it does appear to be worthwhile to constuct examples where there is no such danger.

\begin{figure}[ht]
  \centering \includegraphics[width=3in]{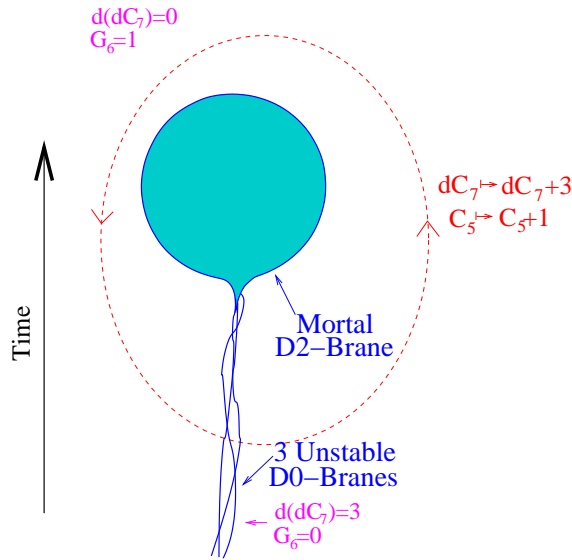}
\caption{Consider massless IIA SUGRA on a spacetime with 3 units of $H$ flux on a 3-cycle, stacks of 3 D0-branes are unstable.  The decay occurs via a mortal D2-brane.  As the D2-brane sweeps out the 3-cycle, it leaves behind it one unit of $G_6$ flux.  The wedge product $G_6\wedge H=k$ is nontorsion and so again the final configuration does not correspond to any twisted K-theory class.} \label{mms2}
\end{figure}

To this end we consider \cite{MMS} the decay of a stack of $k$ D0-branes propagating in the same spacetime as above.  This is depicted in Figure~\ref{mms2}.  Begin with $k$ D0-branes, $k$ units of $H$ flux and no nontrivial $G_6$ flux.  These D0-branes are unstable and may be absorbed by a D2-brane which appears from nothing, sweeps out the 3-sphere and then shrinks until nothing remains except for $G_6$ flux whose integral over the 5-cycle crossed with the noncompact spatial direction yields $1$:
\begin{equation}
\int_{M^5\times\R}G_6=1.
\end{equation}
Again this flux does not correspond to any twisted K-theory class because
\begin{equation}
d_3 G_6=Sq^3 G_6 + H\wedge G_6 = 0 + kx^9=kx^9
\end{equation}
is nonzero.  Here $x^9$ generates the ninth compactly supported cohomology class of a timeslice.   

To better understand the origin of this flux we will describe the field configuration at each step in the decay process explicitly.  In the original configuration there was a stack of $k$ D0-branes.  These are sources for $G_8$ and so
\begin{equation}
ddC_7=dG_8-H\wedge G_6=\delta(\textup{D0-branes})=kx^9. \label{sup}
\end{equation}  
Every timeslice is the product of an 8-cycle and the real line.  The stack of D0's therefore provides a domain wall in a projection onto the real line.  $dC_7=G_8-H\wedge C_5$ may be integrated over the 8-cycle which projects to some point and (\ref{sup}) implies that this integral jumps by $k$ when this cycle is pushed past the stack of D0-branes.  

Further Eq.~(\ref{sup}) implies that this is the only deformation that changes the integral.  In particular, we may consider a 1-parameter family of 8-cycles which is the preimage, under the projection map
\begin{equation}
M^8\times\R^{1,1}\longrightarrow\R^{1,1}
\end{equation}
of a circle linking the mortal D2-brane.  Going around this circle the integral of $G_8-H\wedge C_5$ over the 8-cycle increases by $k$ when the cycles intersects the D0-branes but otherwise does not change.  This means that the cycle enjoys a nontrivial monodromy
\begin{equation}
\int_{M^8} G_8-H\wedge C_5\mapsto k+\int_{M^8} G_8-H\wedge C_5.
\end{equation}
This is consistent with the fact that $G_8-H\wedge C_5$ is valued in $\Z_k$, as predicted by the twisted K-theory classification.  The integral of $G_8$, on the contrary, is well-defined which means that it cannot be subject to any monodromies.  $H$ is held constant everywhere in this section, its integral over the 3-cycle is always equal to $k$.  This means that $C_5$ must be subject to the monodromy
\begin{equation}
\int_{M^5} C_5\mapsto -1+\int_{M^5} C_5.
\end{equation}  
This is true, because the integral of $G_6$ over the corresponding circle-valued family of 5-cycles is equal to $-1$
\begin{equation}
\Delta \int_{M^5} C_5=\int_{S^1\times M^5} G_6=\int_{B^2\times M^5} dG_6=\int_{B^2\times M^5} \delta(\textup{D2-brane})=-1
\end{equation}
because the cycle links the D2-brane.  The fact that $C_5$ is well defined only modulo 1 is consistent with the fact that it is classified by $U(1)$-valued cohomology.  Notice that a shift in $C_5$ by 1 does not change the partition function of a D4-brane to which it couples electrically because the action is exponentiated (and we have absorbed a factor of 2$\pi$ into our measure.)

We have now seen two explanations for the fact that $dC_7$ is well-defined only modulo $k$.  The first is that $G_8=dC_7+H\wedge C_5$ is well defined while $C_5$ is an element of $U(1)$-valued cohomology.  The second is that in the presence of D2-branes, even D2-branes that carry no net D2-brane charge and only exist for a finite period of time, the $\Z$-valued cohomology class of $dC_7$ is subject to a nontrivial monodromy.  In this example the D2-brane in question existed because it described the decay of the D0-branes in the initial condition.  However this initial condition was not necessary for mortal D2's to form, nor was the nontriviality of $dG_8$.  So long as there is a three-cycle supporting $H$-flux (which does not cancel the relevant $W_3$'s) D2-branes may spontaneously appear and sweep out a 3-manifold leaving a stack of D0-branes, which may later decay by the inverse process.  This process will occur frequently if the system is probed at an energy higher than $V_3/g_s$ where $V_3$ is the volume of the 3-cycle.  However it seems unlikely that the well-definedness of the form should know whether such a process is occuring somewhere far away and perhaps not yet causally connected, and so one may expect that at any energy-scale $dC_p$ fluxes are well-defined only modulo $k$, where $k$ is the greatest common divisor of the integrals of the $H$ fluxes over 3-submanifolds of the Poincare dual of $dC_p$ (more precisely $H$ runs over the factors of $dC_p$ in the compactly supported cohomology ring).

This monodromy generalizes the usual monodromy which acts on the fields of IIB string theory in the presence of a D7-brane, which is responsible for the fact that $dC_2$ is well defined only modulo $H=k$.  To see this, compactify the spatial real line above, T-dualize and decompactify.  Of course this compactification must be done without the net D-brane charge of the above examples, as the flux would have no where to go.  However if after the T-duality to IIB we decompactify the circle then we may consider the T-dual of the process described in the previous subsection.  Begin with $k$ D5-branes and no $G_1$-flux.  These D5-branes may then decay via the nucleation, wrapping and destruction of a mortal D7-brane.  The integral of $G_1$ over a circle which links the D7-brane once is $-1$.  This means that $C_0$, the IIB axion, has a monodromy of 1.  A 3-sphere wrapping the nontrivial 3-cycle may be dragged around the loop and the integral of $G_3$ over this sphere will incriment by $k$ as it crosses the stack of D5-branes.  This is the usual monodromy
\begin{equation}
H\mapsto H\sp G_3\mapsto G_3+H
\end{equation}
associated with loops around a D7-brane, but in this case it is the same ill-definedness that leads to the $Z_k$ classification of $G_3$ in the twisted K-theory framework.

\section{$E_8$ Interpretation} \label{e8sec}

We have seen that twisted K-theory appears to classify the configurations of fields in the absence of D-branes but does not discern configurations containing fields $dG_{p+1}$ which are nontrivial only compactly-supported cohomology and vanish in the usual cohomology.  Such fields, of course, can only exist on noncompact manifolds.  To begin to understand why such configurations are missed by the twisted K-theory classification we consider a more powerful classification scheme, which incorporates all of the data of the NS and RR fields of IIA.  This is the $E_8$ gauge bundle formalism \cite{FluxQuant,DMW,Madrid,Allan}, which is conjectured to contain within it the twisted K-theory classifications of branes ignoring fluxes and fluxes in the absence of branes as special cases \cite{Horava,Allan}.  Included here is a preview of how these classifications are realized which leads to a justification for the absence of the D-brane decay residues in twisted K-theory.

\subsection{$E_8$ in M-Theory} 

``$E_8$ gauge bundle formalism'' means that we begin with some 11-dimensional M-theory configuration and for each $G_4$ four-form field strength we create an $E_8$ gauge bundle such that $G_4$ is roughly its instanton charge, more precisely
\begin{equation}
\frac{G_4}{2\pi}=\frac{\textup{Tr}(F^2)}{16\pi^2}-\frac{\textup{Tr}(R^2)}{32\pi^2}. \label{cons}
\end{equation}
Due to the uniquely simple topology of $E_8$, the fact that $\pi_3(E_8)=\Z$ is the only nonvanishing low-dimensional homotopy class, $E_8$ is the only group for which such a condition uniquely specifies the topology of the bundle.  An important construction is the nontrivial $E_8$ bundle over the four-sphere.  The subbundles over the northern and southern hemispheres are trivialized and the transition function is a map from the equatorial $S^3$ to $E_8$, which is a nontrivial element of $\pi_3(E_8)$.

The soliton spectrum of M-theory on a topologically trivial patch of spacetime is easily reproduced.  The M5-brane is the defect such that any $S^4$ has the nontrivial $E_8$ bundle over it whose transition function represents the class in $\pi_3(E_8)$ equal to the linking number of the $S^4$ and the M5-brane.  In our topologically trivial patch of space we can introduce Minkowski coordinates.  We can then introduce an M5-brane and ignore the fact that it has an infinite throat by cutting out a tubular neighborhood, but leaving the nontrivial bundles that link it behind.  Really we only need to cut out the M5-brane itslf, which was already infinitely far away from everything else, and we can reparametrize the coordinates to leave 11-dimensional Minkowski space minus a 6-manifold. 

\begin{figure}[ht]
  \centering \includegraphics[width=3in]{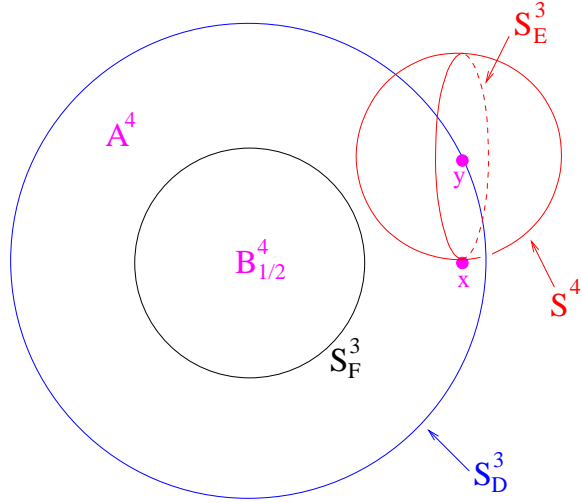}
\caption{An M5-brane wraps a trivial 3-cycle $S^3_D$.  This leads to a nontrivial $E_8$ bundle over every linking $S^4$, characterized by a transition function $S^3_E\longrightarrow E_8$.  This transition function represents a nontrivial class in $\pi_3(E_8)$.  A choice of $x\in S^3_E$ in each $S^4$ is mapped to some basepoint in $E_8$ for each $S^4$.  $S^4$'s are labeled by their centers $y\in S^3_D$, and thus for each $x$ there is a map $S^3_D\longrightarrow E_8$.  $E_8$ approximates the classifying space $K(\Z,3)$ and therefore the homotopy class of the map determines an element of the 3rd cohomology group of the M5-brane's worldvolume.  This element is the worldvolume 3-form fieldstrength $T_3$ and so the homotopy class of this map equals the M2-brane charge of the dielectric M5-brane.} \label{e8}
\end{figure}

Now we can make an M2-brane using the Myers dielectric effect.  Consider an M5-brane whose intersection with this patch is $\R^{1,2}\times S^3$.  To avoid confusion we will name this 3-sphere $S^3_D$.  We will consider a configuration in which everything is constant over the $\R^{1,2}$ along which the M5 extends and we will project it out.  Our coordinate patch is trivial, and so the $S^3_D$ is contractible and the configuration carries no net M5 charge.  The configuration does however carry M2 charge which is equal to
\begin{equation}
Q_{M2}=\int_{S^3_D}T_3 \label{m2}
\end{equation}
where $T_3$ is the self-dual 3-form field strength which propagates on the M5-brane worldvolume.  To interpret $T_3$ we will need to study the topology of the $E_8$ bundle over $\R^8 - S^3_D$.  

Choose $S^3_D$ to be the unit 3-sphere along an $\R^4$ plane in $\R^8$.  Consider a small, round 4-sphere centered at some point $y\in S^3_D$.  This 4-sphere extends in the radial direction of $S^3_D$, and we will call its point of closest approach to center of $S^3_D$ the 4-sphere's north pole.  The 4-sphere's equatorial 3-sphere, which we will call $S^3_E$, is a sphere of fixed radius in the $\R^4\subset\R^8$ orthogonal to $S^3_D$.  Trivialize the $E_8$ bundle over the northern and southern hemisphere of the 4-sphere.  The 4-sphere links a (deleted) M5-brane and so the transition function is a map
\begin{equation}
f_y:S^3_E\longrightarrow E_8\sp [f]=1\in\pi_3(E_8)=\Z.
\end{equation}  
Notice that $f$ is the transition function for a given 4-sphere, but the function and in particular its basepoint may depend on which 4-sphere is chosen.  There is one 4-sphere for each point $y\in S^3_D$, the 4-sphere whose center is $y$.  Thus there is a family of $f$'s, one for each $y$
\begin{equation}
F:S^3_D\times S^3_E\longrightarrow E_8.
\end{equation}
In particular if we choose a point $x\in S^3_E$ then we may define a new function
\begin{equation}
g:S^3_D\longrightarrow E_8:y\mapsto F(y,x)=f_y(x)\sp [g]\in\pi_3(E_8)=\Z.
\end{equation}
Thus the choice of $E_8$ bundle over this space is not determined by the shape of the M5-brane alone, but also by the homotopy class of the function $g$. 

This homotopy class is precisely the M2-brane charge of the configuration.  To see this, consider the following situation.  Set the homotopy class to be zero and the integral of $G_4$ over the 4-ball bounded by $S^3_D$ to be $k$:
\begin{equation}
[g]=0\sp \int_{B^4}G_4=k\sp \partial B^4=S^3_D.
\end{equation}
This means that the $E_8$ bundle over $B^4$ has an instanton number of $k$.  We can divide $B^4$, which has a radius of one, into two patches: an inner patch $B^4_{1/2}$ which is a $4$-ball of radius $1/2$ and the rest of $B^4$ which is an annulus $A^4$.  The $E_8$ bundle may be trivialized over these two patches, with a transition function $h$ over the boundary 3-sphere, called $S^3_F=\partial B^4_{1/2}$, which represents the homotopy class $k$
\begin{equation}
h:S^3_F\longrightarrow E_8\sp [h]=k\in \pi_3(E_8)=\Z.
\end{equation}  
Our $S^3_D$ lives in the patch $A^4$, but we may contract it, bringing it across $S^3_F$, into the patch $B^4_{1/2}$ where the bundle is trivial.  This changes the transition function on $S^3_D\times S^3_E$ by composing it with $h$
\begin{equation}
F(x,y)\mapsto F\p (x,y)=h(x)F(x,y).
\end{equation}
In particular this changes the homotopy class of the transition function $g$
\begin{equation}
g(x)\mapsto g\p (x)=h(x)g(x)\sp
[g]\mapsto [h]+[g]=k+0=k.
\end{equation}
Thus the homotopy class $[g]$ shifts by $k$ as the M5-brane sweeps out $k$ units of $G_4$ flux, or equivalently as the $C_3$ flux on its worldvolume shifts by $k$.  

$C_3$ is not a gauge-invariant quantity, in the worldvolume of an M5-brane the gauge-invariant quantity is $C_3+T_3$.  On the topologically trivial patch $B^4_{1/2}$ we would like to eliminate $C_3$ so that we may shrink the M5-brane to nothing and see if it disappears, but gauging away $C_3$ leaves us with $k$ units of $T_3$, which are the desired $k$ units of M2-brane charge according to Eq.~(\ref{m2}).  Of course we may also see from the topology of the bundle that this configuration is nontrivial and cannot be contracted away.

To see how the torsion classes of K-theory show up is now not difficult.  The simplest approach would be to notice that we can make a 4-cycle with $k$ units of $G_4$ flux by simply identifying $S^3_D$ with a point in the above example.  Then the $k$ units of membrane charge that were found would be trivial, as the original M5-brane on $S^3_D$ contained no M2 charge and wrapped no volume and so would simply disappear, whereas a stack of k-membranes could grow into an M5-brane wrapping a trivial $S^3$ in $S^4_{1/2}$ and continue to grow until it wrapped $S^3_D$, at which point the M2 charge is gone and the branes have disappeared.  Thus M2-branes in this background are classified by $\Z_k$.  Replacing the 4-sphere with $S^3\times S^1$ and dimensionally reducing on the $S^1$ we find $k$ units of $H$ flux and the $\Z_k$ above reduces to the $\Z_k$ of twisted K-theory.

If we think of $g$ as an inclusion of the M5-brane into the total space of the $E_8$ bundle then {\textbf{M2-branes are classified by the homology of the total space of the bundle.}}  The homology of the total space $E$ of the bundle in this case contains the necessary $H_3(E)=\Z_k$.  This picture is appealing because $T_3$ has the simple interpretation as the Jacobian determinant of the embedding map\footnote{The determinant of the $3\times 3$ submatrix specified by the indices on $T_3$.}, which suggests that perhaps the bosonic part of the M2-brane action has a simple 259d interpretation such as the volume of the image of this embedding with the $T_3$ terms in its worldvolume action coming from the dimensional reduction of the $E_8$ fibres.  Notice that in this perspective the M2-brane is just an M5-brane (a 6-manifold in $E$) that wraps the 3-sphere in the group manifold $E_8$ and also extends in 3 directions along the 11-dimensional base space.

\subsection{$E_8$ in Type II String Theory}
We are actually interested in type II string theory and not M-theory.  Following the suggestion of \cite{Horava}, these ideas extend to type IIA using the isomorphism of $E_8$ bundles over circle bundles over a 10-manifold $M^{10}$ to $LE_8$ bundles over $M^{10}$, where $LE_8$ is the trivially centrally-extended free loop group of $E_8$ \cite{Slow}.  Topologically this loop group is a circle crossed with the group manifold $E_8$ crossed with the based loop group of $E_8$.  The circle is the M-theory circle.  Now instead of maps from $g$ to $E_8$ we have maps to $LE_8$, which decompose into maps to a circle, maps to the group manifold $E_8$ and maps to the based loop group of $E_8$.  The fundamental and adjoint representations of $E_8$ are the same, and so there is an isomorphism between maps to the $E_8$ group manifold and sections of an associated vector bundle, which we may then use to construct sections of the vector bundle associated to the loopgroup.  

The based loop group approximates $K(\Z,2)$, it only differs in homotopy groups of dimension at least 14, which appear to be irrelevant here. This suggests that Proposition 7.2 of Ref.~\cite{Bgerbe} may then be applied to demonstrate that topologically distinct bundles are in a one to one correspondence with elements of twisted K-theory.  The condition that these bundles be free of defects is the condition that they contain no D-branes.  Similarly the defects themselves, the D-branes, should be classified by twisted K-theory.  To prove such a correspondence, one would need to show, among other things, that cohomology classes that do not lift to K-theory correspond to obstructions to creating the $LE_8$ bundle.  In the case of D6-branes this has already been shown by Sethi \cite{SethiD6}, in particular this anomaly is an obstruction to the existence of the $S^1$ piece of the bundle.  The D8-brane anomaly was studied in Ref.~\cite{Slow} with the conclusion that the conditions, originally derived from SUGRA, of how many branes must end on each type of brane in massive IIA are precisely the conditions under which the $LE_8$ bundle exists.  In the case of D6 branes ending on D8-branes this is the Freed-Witten anomaly, which is responsible for solitons being classified by K-theory instead of cohomology.

One also needs to show that cohomology classes which are identified by the K-theory classification correspond to topologically equivalent $LE_8$ bundles.  In the $spin^c$ case this identification relates fluxes which differ by the product of $H$ and another cohomology class.  For simplicity, consider a non-contractible 3-sphere that supports K-units of $H$ flux
\begin{equation}
\int_{S^3} H=k.
\end{equation}
It is important that the sphere is noncontractible, because otherwise there would be NS5-branes and we would not expect the twisted K-theory classification to apply.  We have seen above that in M-theory this yields a $S^1\times S^3$ which supports $k$ units of $G_4$ flux, and so the $E_8$ bundle has an instanton number of $k$.  We may trivialize the bundle over the northern and southern hemispheres of the $S^3$, and then the transition function is a map from its equator crossed with $S^1$ to $E_8$, that is a map
\begin{equation}
S^2\times S^1\longrightarrow E_8
\end{equation}
which is equivalent to a map from $S^2$ to $LE_8$.  Thus in IIA our 3-sphere is the base of a nontrivial $LE_8$ bundle with a transition function on its two-sphere equator that represents the class $k$ in $\pi_2(LE_8)=\Z$.  The two-sphere in $LE_8$ generates $H_2(LE_8)=\Z$ but in the total space $E$ of the bundle this is a torsion class $H_2(E)=\Z_k$. 

Now we may try to classify D$p$-branes that do not wrap this 3-sphere.  Equivalently, using the above technology, we may classifiy D$(p+2)$-branes that carry D$p$-brane charge because they wrap the 2-sphere in $LE_8$.  This 2-sphere generates a $k$-torsion homology class in the total space, and so these D$p$-branes are classified by $k$-torsion.  This is precisely the identification made by the twisted K-theory classification.  And so, at least in this class of examples, the twisted K-theory identifications are contained in the topology of the $LE_8$ bundle.  In the $spin^c$ case the restriction on possible branes also appears to be manifest in the $LE_8$-formalism, because a D$p$-brane that wraps a 3-cycle supporting $H$-flux may be interpreted as a D$(p+2)$-brane which wraps the 2-cycle in the fiber, but this 2-cycle cannot be extended everywhere.  In particular, it may be extended everywhere except for $k$ Dirac strings.  These Dirac strings, in the case of D8-branes, have been shown \cite{Slow} to be precisely the $k$ D$(p-2)$-branes that must end on the D$p$-brane to cancel the Freed-Witten anomaly.  These two types of evidence suggest that choice of twisted K-theory class depends on the topology of the $E_8$ bundle, although a no precise correspondence has been made in the general case.



The above construction of course only applies to type IIA, and no analogous formulation is known for IIB on a general spacetime.  However if type IIB is compactified on a 10-manifold which is a circle fibred over a 9-manifold then we may proceed\footnote{U. Varadarajan independently reached conclusions similar to those that follow.}.  We compactify IIA on a circle and use the loop group correspondence again.  This time we get a 9-manifold with an $LLE_8$ bundle fibered over it.  
\begin{equation}
\pi_1(LLE_8)=\Z^3
\end{equation}
there are two circles from the central extension after each compactification, which are just the two circles that M-theory was compactified on.  We will call these $S^1_M$ and $S^1_{IIA}$.  There is a third circle which is the nontrivial circle in the based loop group of the based loop group of $E_8$, that is it decends from $\pi_3(E_8)$.  We will call this circle $S^1_{E8}$.  

The electrically charged objects corresponding to $\pi_3(E_8)$ are M2-branes and so the electrically charged objects under $S^1_{E8}$ are membranes which wrap the other two circles.  These are fundamental strings wrapping $S^1_{IIA}$ and so have a mass of $R_{IIA}/\alpha\p$.  This is the mass of the lowest KK mode after a reduction on $S^1_{E8}$ and so the radius of $S^1_{E8}$ is
\begin{equation}
R_{E8}=\frac{1}{R_{\textup{IIA}}/\alpha\p}=\frac{\alpha\p}{R_{\textup{IIA}}}.
\end{equation}
The KK reduction is invalid when $R_{IIA}$ is smaller than $R_{E8}$, which happens when $R_{IIA}$ is smaller than $\sqrt{\alpha\p}$, but in this case we may KK reduce $S^1_{IIA}$ and leave $S^1_{E8}$ as an ``external'' direction.  This leaves the other two circles as an internal torus, known in the literature as the F-theory torus \cite{Ftheory}.  Although we have swapped two circles, we have not changed the topology of the total space and so we may arrive at a K-theory classification, although of course dimensions of forms are shifted by one because an internal and external circle has been swapped.  In fact we want this shift in dimensions, because we want fields to be classified by $K^1$ instead of $K^0$, which differs exactly by a wedge product with a new circle.  

There are many simple checks on this identification which may be done.  For instance, one may show that an NS5-brane in IIA which does not wrap $S^1_{IIA}$ is T-dual to a KK-monopole in IIB.  Also a D6-brane which does not wrap is dual to a D7-brane with the usual F-theory interpretation.  T-dualizing the D7-brane again about a circle which it does not wrap yields IIA once again, now with an $LE_8$ apparently unrelated to the original $LE_8$ fibers, with a D8-brane that has a nontrivial circle fibration over the $K(\Z,2)$ fibers.  This confirms a conjecture concerning the $E_8$ interpretation of massive IIA made in \cite{Allan} and further tested in \cite{Slow}, but the above approach has the advantage that the T-duality to Hull's proposal \cite{Hull} may be seen explicitly.  We have done these checks and they will appear elsewhere.

This was a long tangent, but finally we may address the relevant question to this paper, which is why D-brane decay reminants do not appear to correspond to twisted K-theory classes.  First we note that we may start in a configuration with no D-branes that is classified by twisted K-theory, the processes described above may then create D-branes which later decay into nothing.  The final state here is the same as the initial state and is described by twisted K-theory, despite the fact that D-branes have decayed.  The reason that these D-branes, unlike those of the previous section, are captured by the K-theory classification is that these do not support $dG_{p+1}$ flux.  These carry D-brane charge because they violate the Bianchi identity (\ref{bianchi}), but instead of violating the Bianchi identity with a $dG_{p+1}$ term they violate it with a nontrivial $H\wedge G_{p-1}$.  The fact that $H\wedge G_{p-1}$ is supported on the unstable brane means that twisted K-theory does not capture the flux on the unstable brane, but the twisted K-theory classification is only supposed to apply to backgrounds with no D-brane charges so this poses no problem.  When the unstable branes decay the undesired flux is promptly canceled by the flux emitted by the mortal brane responsible for the decay.  This leaves the original flux configuration, which was given to be described by twisted K-theory.  This $H\wedge G_{p-1}$ which was created and violated the Bianchi identity temporarily may be understood in terms of the topology of the $LE_8$ bundle just as the $G_4\wedge G_4$ flux sourced by an M2-brane was understood in the last subsection.  The only difference is that the dimensions of the spheres, balls, annuli and forms have changed.  

By contrast, the unstable D-branes of the previous section were Poincare dual to $dG_{p+1}$ while $H\wedge G_{p-1}$ vanished.  The case of unstable D2-branes and a mortal D4 lifts to a picture similar to our M-theory picture.  Yet this configuration is not quite the same.  Instead of nontrivial $G_4\wedge G_4$ flux, we have nontrivial $d*G_4$ flux.  After the unstable membranes have decayed the Bianchi identity must be solved, which yields a nontrivial $G_4\wedge G_4$ flux, but we still will have the $d*G_4$ flux.  This expression contains a Hodge star and thus is not a topological quantity in our $E_8$ bundle.  In fact, it is possible that topologically equivalent $E_8$ bundles correspond to distinct $d*G_4$ fluxes.  A geometrical description of the $E_8$ bundle will of course contain the information about this flux, as the flux is determined by $G_4$ and $G_4$ is determined from the data of the bundle via Eq.~(\ref{cons}).  However the twisted K-theory class is determined by the topology of the $E_8$ bundle, and the topology of the $E_8$ bundle does not appear to be sensitive to the presence of such D-brane decay reminants.

\section{Mortal NS5-Branes} \label{ns5sec}
Twisted K-theory is only an approximate classification of RR field configurations, reflecting an approximation which is thought to be valid under some sort of weak-coupling condition\footnote{However the success of the analogous $E_8$-bundle classification in M-theory (which has no weak-coupling regime) as well as the fact that D-brane charges are conserved even if the background $H$ fluxes temporarily change throughout the universe may lead one to believe that such a condition may be relaxed with a suitable generalization of K-theory.}.  The nature of this approximation has never been made precise and will not be made precise here, rather we will use the phenomena described above to understand how the classification of fields by K-theory twisted by $H$ breaks down when NS fluxes are no longer held constant.  Here we are only concerned with a topological description of the fluxes, therefore allowing $H$ and $*H$ to vary is equivalent to including NS5-branes and fundamental strings.  The fact that treating NS fluxes on equal footing with RR fluxes must lead to a breakdown of the twisted K-theory classification was seen already in, for example, Ref.~\cite{DMW} which demonstrated the incompatibility of the K-theory classification with S-duality in IIB.

\begin{figure}[ht]
  \centering \includegraphics[width=3in]{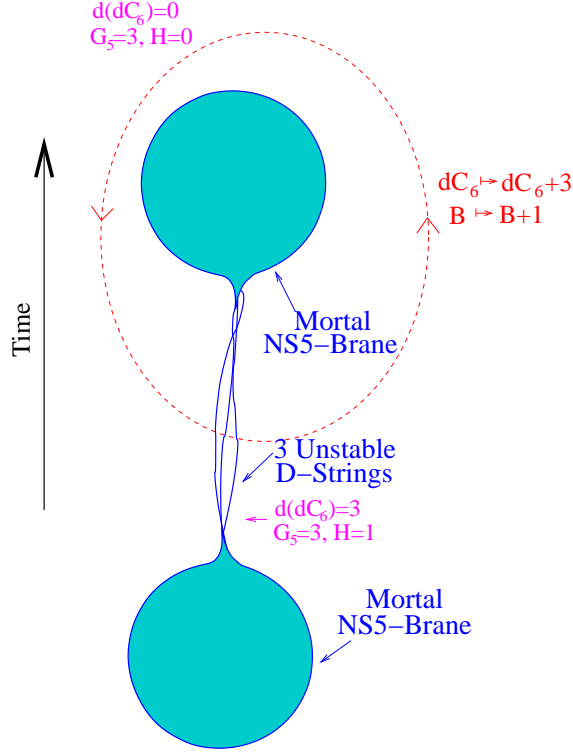}
\caption{An NS5-brane sweeps out a 5-cycle which supports 3 units of $G_5$ flux.  When it decays, it leaves 3 D-strings.  Later these D-strings decay via the same process by which they were created.  The initial and final states are included in the twisted K-theory classification.  However the monodromy about the NS5-brane relates distinct cohomology classes which do \textit{not} represent the same K-theory class.  This is one way in which the K-theory classification is modified by the inclusion of processes involving NS5-branes.}
\label{ns5a}
\end{figure}

There are two differences between the set of elements of twisted K-theory and those of integer valued cohomology, which provide the physical motivation for classifications of fluxes by K-theory instead of cohomology.  The first is that many elements of cohomology do not correspond to elements of K-theory.  We have seen that, in the $spin^c$ case, this appears to have the physical interpretation of excluding configurations which contain the decay products of nontrivially-wrapped D-branes.  The second is that many distinct elements of cohomology correspond to the same K-theory class.  This was physically interpreted as the fact that fields are defined only up to certain monodromies corresponding to translation around possible mortal branes.  Whether these mortal branes exists or not should not affect physics far away, and so fluxes are classified only up to the actions of the monodromies that would exist were there to be such mortal branes somewhere.  Of course there are processes that are neglected in this reduction from cohomology to K-theory.  Such processes are the subject of the present section.  

One omission is that monodromies about NS5-branes are not considered.  For example, consider a spacetime with a nontrivial 3-cycle $Z^3$ that supports $j$ units of $G_3$ flux.  Begin with no other fluxes and no branes.  The spacetime may be compact or noncompact.  An NS5-brane may spontaneously form which during its lifetime sweeps out $Z^3$ and is extended along 3 irrelevant directions.  When the NS5-brane tries to shrink to a point on the other side of $Z^3$ it now carries $j$ units of D3-brane charge.  This leaves a stack of $j$ D3-branes, which later may decay by the inverse process of their creation.  The D3-branes are sources for $dC_4$ flux, and so the lack of conservation of D3-brane number appears to imply that a mondromy around the NS5-brane shifts the cohomology class of $dC_4$ by $j$ units.  Alternately this may be seen by noting that $G_5$ is gauge invariant and so should be well defined, while $B$ shifts by one unit.  An application of the Bianchi identity (\ref{bianchi}) then yields the desired monodromy.
\begin{equation}
G_5=dC_4-B\wedge G_3 \mapsto G_5\sp
G_3=j\mapsto G_3\sp
B\mapsto B+1\sp
dC_4\mapsto dC_4+j.
\end{equation}
Similarly we could have considered a background $G_5$ flux and unstable D-strings, as in Figure~\ref{ns5a}, which would have led to a monodromy for $dC_6$
\begin{equation}
G_7=dC_6-B\wedge G_5 \mapsto G_7\sp
G_5=j\mapsto G_5\sp
B\mapsto B+1\sp
dC_6\mapsto dC_6+j.
\end{equation}
These relations are of course the flux versions of the known terms in the S-duality covariant AHSS for charges.  An application of this viewpoint to a non-$spin^c$ example may be a useful step in the formulation of an S-duality covariant AHSS for fluxes.  Notice that in the $spin^c$ case these arguments extend quite naturally to IIA, where we may also consider D2-branes and D4-branes whose nonconservation leads to mondromies for $dC_5$ and $dC_3$ fluxes respectively. 

One may now wonder whether there exists an approximation in which it was valid to omit these equivalences.  Perhaps there is some weak-coupling limit which reproduces this approximation, for example if one considers a box in spacetime of size $O(1/g_s)$ then one may expect to find some mortal D-branes but NS5-branes are much heavier and so would be unlikely to appear at small coupling.  The static space-time topology could then only be recovered in the weak-coupling limit, where the box size and thus the extent of the time coordinates become infinite.  

Another missing effect is that while fluxes resulting from D-brane decay are explicitly dropped, there is no such treatment of the decay products of unstable F-strings.  At least in the compact case such an exclusion is necessary to assure that F-strings do not spontaneously form if their fluxes would have no place to go.  More explicitly, consider an initial condition consisting of a $spin^c$ spacetime with $j$ units of background $G_p$ flux on some $p$-cycle, $j$ fundamental strings which perhaps wrap a circle or perhaps extend to infinity, and no other fluxes or branes.  Now the $j$ fundamental strings may decay by blowing up into a dielectric Dp-brane which sweeps out the $p$-cycle and then disappears, while also extending in whatever other direction was spanned by the F-strings.  

For concreteness we now consider the case $p=3$.  The F-strings are a source for $d*H$ flux, more precisely their charge is defined by the violation of the equation of motion
\begin{equation}
G_3\wedge G_5=d*H.
\end{equation}
$G_5$ is given to vanish in the initial condition and so the F-strings must source the dual NS flux.  When the strings decay the Bianchi identity is restored because the D3-brane creates a $G_5$ flux.  However this leaves
\begin{equation}
G_3 \wedge G_5 \neq 0
\end{equation}
which is S-dual to a $G_5$ flux which is not annihilated by the AHSS differential $d_3$.  Therefore this flux configuration is not part of a classification scheme which extends twisted K-theory by ``imposing'' S-duality.  Reversing the logic, an S-duality covariant extension of K-theory should not capture such configurations.  Now we may ask why we were allowed to ignore these in our approximation.  It is a bit more difficult to see why this effect should go away at small $g_s$, where presumably F-strings become much more plentiful than D-branes.  But whether this is a reasonable approximation at weak coupling requires an understanding of what it means to omit such fluxes in the first place.  

\begin{figure}[ht]
  \centering \includegraphics[width=3in]{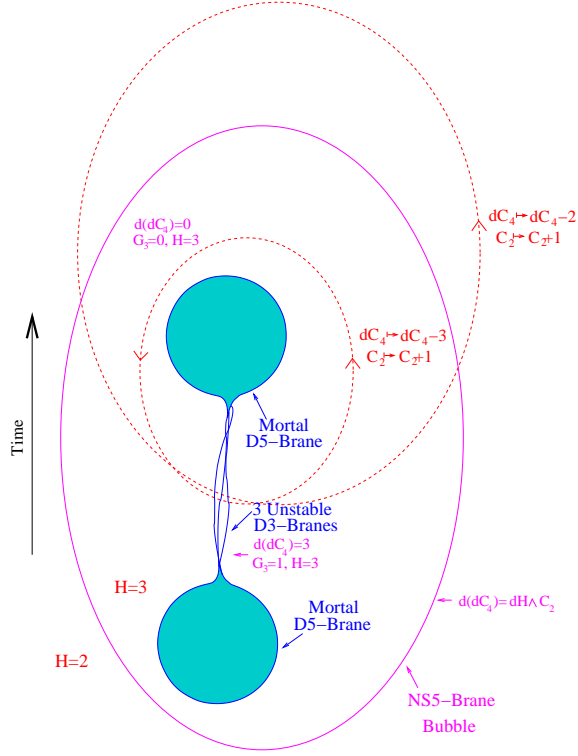}
\caption{In a background with two units of $H$ flux a mortal D5-branes sweeps out a 3-cycle, leading to the creation of unstable D3-branes which later decay by the same process.  The initial and final states are described by twisted K-theory and this is the kind of process which leads to an identification of $dC_4$ fluxes that differ by two units.  However in this case the entire process is linked by an NS5-brane bubble, which leads to an additional unit of $H$ flux inside the bubble.  This means that fields inside the bubble are identified modulo $3$, while those outside continue to be identified modulo $2$.  Alternately this may be read from the monodromies, observing that $ddC_4$ is supported on the NS5-brane bubble, and its support depends on $C_2$ which itself experiences a monodromy and thus its value is one greater as the loop leaves the bubble than when the loop entered, leading to the expected one unit shift in the $dC_4$ monodromy.}
\label{ns5c}
\end{figure}

A third missing process, drawn in Figure~\ref{ns5c}, is a topologically trivial bubble of NS5-branes which forms and decays in spacetime.  The result is a region of spacetime with a different $H$ flux from the ambient spacetime.  Any of the above processes may occur inside of the bubble, where ``inside'' means that the 3-cycles wrapped by mortal branes link the NS5-brane and so the $H$ flux integrates to $k+1$ instead of $k$.  Inside of the bubble $dC_p$ fluxes are therefore conserved modulo $k+1$ instead of modulo $k$, as a result of the monodromy caused by a mortal D$(9-p)$-brane
\begin{equation}
C_{p-2}\mapsto C_{p-2}+1\sp
dC_p=G_{p+1}-H\wedge C_{p-2}\mapsto G_{p+1}-H\wedge (C_{p-2}+1)=dC_p-(k+1).
\end{equation}
However outside of the bubble there are still only $k$ units of $H$ flux, and so the $C_{p-2}$ monodromy of one unit only leads to a $dC_p$ monodromy of $k$ units outside of the bubble.  This means that if there are two regions of spacetime with different amounts of H flux, the RR fluxes in each region will be classified by the corresponding twisted K-theory.  One must bear in mind of course that NS5-branes, being of codimension more than 1, do not actually split spacetime into multiple regions.  Thus the notion of a region needs to be interpreted very carefully, in terms of how many times a given 3-cycle links the NS5-brane.  In fact this implies that there are an infinite number of ``regions'', but BPS configurations are unlikely to exist in more than two.


\section{Conclusion}

The twisted K-theory classification of fields implies that distinct cohomology classes must be somehow physically equivalent.  We have seen that this physical equivalence results from monodromies generalizing the well-known physical equivalence of fields in the presence of a D7-brane.  These monodromies are caused by the mortal D-branes \cite{MMS} that are responsible for the the decay of unstable D-branes that wrap non-trivial cycles.  When the zero-coupling restriction is relaxed mortal NS5-branes also must be considered, and these lead to further identifications which modify the twisted K-theory classification.  Such corrections may be expected, for example, from the lack of S-duality covariance of the twisted K-theory classification of fluxes in type IIB \cite{DMW}.  

The final states of brane-decay processes provide examples of fluxes which are nontrivial in compactly supported cohomology but trivial in ordinary cohomology.  The usual K-theoretic classification is not sensitive to such fluxes, and in fact the notion of field strength which does include these fluxes does not respect the forementioned equivalences of configurations related by brane decays.  In fact this notion of field strength is not even quantized.

To gain an alternate understanding of the failure of twisted K-theory to capture compactly-supported fluxes, we previewed the relation between the K-theory classification and the $E_8$ bundle formalism.  The $E_8$ bundle formalism appears to reproduce twisted K-theory as a set of sections constructed explicitly above in the M-theory case.  
The compactly-supported cohomology classes in question did not appear to affect the topology of the $E_8$ bundle, although they did affect the geometry.  As the twisted K-theory class of a configuration was determined by the topology of the corresponding bundle, twisted K-theory could not be sensitive to these classes.  Many details in this construction were left to work in progress. 

\noindent 
{\bf Acknowledgements}

\noindent
Jarah would like to thank himself for many discussions, some of which he found useful.
Jarah has been brought to you by the INFN Sezione di Pisa.

\noindent

\bibliographystyle{ieeetr} 
\bibliography{twist}
\end{document}